\def\be{\begin{equation}}
\def\ee{\end{equation}}
\def\ba{\begin{array}}
\def\ea{\end{array}}

\documentclass[groupedaddress,aps,superscriptaddress,showpacs,nofootinbib]{revtex4}%
\usepackage{epsfig,dsfont,amssymb,amsmath,amsthm,amsfonts,amsbsy,mathrsfs}
\usepackage{graphicx, color}
\usepackage{amsmath}
\usepackage{amssymb}
\usepackage{mathdots}
\usepackage{float}
\usepackage{cases}
\usepackage{graphpap}
\usepackage{wrapfig}
\usepackage{tikz}
\usepackage{indentfirst}
\usepackage{picinpar,graphicx}
\begin{document}
\title{Uniform Entanglement Frames}
\author{Yunlong Xiao}
\affiliation{School of Mathematics, South China University of Technology, Guangzhou, Guangdong 510640, China}
\affiliation{Max Planck Institute for Mathematics in the Sciences, 04103 Leipzig, Germany}
\author{Naihuan Jing\footnote{Corresponding Author: jing@ncsu.edu}}
\affiliation{School of Mathematics, South China University of Technology, Guangzhou, Guangdong 510640, China}
\affiliation{Department of Mathematics, North Carolina State University, Raleigh, NC 27695, USA}
\author{Xianqing Li-Jost}
\affiliation{Max Planck Institute for Mathematics in the Sciences, 04103 Leipzig, Germany}
\author{Shao-Ming Fei}
\affiliation{Max Planck Institute for Mathematics in the Sciences, 04103 Leipzig, Germany}
\affiliation{School of Mathematical Sciences, Capital Normal University, Beijing 100048, China}

\begin{abstract}
We present several criteria for genuine multipartite entanglement from universal uncertainty relations based on majorization theory. Under non-negative Schur-concave functions, the vector-type uncertainty relation generates a
family of infinitely many detectors to check genuine multipartite entanglement. We also
introduce the concept of $k$-separable circles via geometric distance for probability vectors, which
include at most $(k-1)$-separable states.
The entanglement witness is also generalized to a universal entanglement witness which
is able to detect the $k$-separable states more accurately.
\end{abstract}


\maketitle
\section{Introduction} 

Entanglement is one of the key characteristics of quantum theory and has been a source for new computational methods
and algorithms in quantum information theory. Numerous criteria of entanglement have been discovered for pure and mixed quantum states, and many of them are devoted to entanglement of bipartite states. For a multipartite system, separability can be classified into $k$-separability \cite{GaH}
and the quantum state is called genuinely entangled
if it is not separable with respect to any tensor bipartition of its space (see Sect. III C for
detailed definition).
Due to the importance of genuine entanglement, numerous approaches have been devoted to detecting genuinely multipartite entanglement. Among the influential ones, the methods using entanglement witness \cite{HuS, dVH, Huber, WuK, HuP, SpV, JuM, MaL}, generalized concurrence \cite{MaC, ChM, HoG, GaY, LiF}, and Bell-inequality \cite{BaG} are very useful.
Nevertheless, the problem of detecting genuinely entanglement is far from being solved.

The principle behind the entanglement witnesses is that entanglement of multiparity
quantum states gives rise to nonlocal correlations of measurement observers, whose measurement outcomes obey certain bounds. In particular
the lower bound of the uncertainty relation is expected to provide better entanglement witnesses.
In this paper, we use the universal uncertainty relation and majorization theory to derive further
criteria for $k$-separability and genuine entanglement.

We take mixed three-qubit states as an example to explain our approach to entanglement using majorization.
In \cite{Acin}, mixed three-qubit states have been classified into three classes of
genuine three-qubit entanglement (W-type, GHZ-type), biseparable and fully separable states.
Our approach not only gives a new method to detect entanglement, but also provides a fine tuned devise to
further classify biseparable states into three subclasses of $AB-C$, $A-BC$, and $AC-B$
types. Furthermore, we introduce the notion of
a {\it universal entanglement witness} and derive its canonical form, which can be used to picture
layers of $k$-separability \cite{GaH} as concentric circles.

The principle behind our majorization method is to use uncertainty relations to detect entanglement.
In the original
Heisenberg-Robertson \cite{Heisenberg, Robertson} uncertainty principle, the lower bound
of the product of standard deviations of two incompatible observables $A$ and $B$ is given by
\begin{equation}
\Delta A\cdot\Delta B\geqslant\frac{1}{2}|\langle[A, B]\rangle|,
\end{equation}
where $\Delta A$ is the standard deviation of $A$ respect to the quantum state $\varphi$. Usually the
uncertainty relations in terms of standard deviations is state-dependent (but see \cite{XJLF}).

Entropy has been used as a measurement \cite{Hirschman} for the uncertainty principle.
Deutsch \cite{Deutsch} showed that the lower bound of the entropy uncertainty relation (EUR) in a finite dimensional Hilbert space is
\begin{equation}\label{e:H1}
H(p)+H(q)\geqslant -2\ln C,
\end{equation}
where $C=(1+\sqrt{c_{1}})/2$ and $c_{1}=max_{i,j}|\langle a_{i}|b_{j}\rangle|^{2}$ is the maximal overlap between the bases ${|a_{i}\rangle}$ and ${|b_{j}\rangle}$, $p$ and $q$ are the probability distributions in the usual manner.
In this form the lower bound is independent on the state. Other uncertainty relations and improvements
have been given in \cite{Maassen} and \cite{Coles}, where the lower bound are mostly state-independent
and computable from two probability vectors.
Good surveys for these bounds
can be found in \cite{Wehner, Blankenbecler, Damgaard, DiVincenzo, Oppenheim, Goehne, Loock, Coless}.

The universality of the
information-based uncertainty principle gives a unbreakable lower bound that holds in general. In fact, the uncertainty relations can be quantified by majorization \cite{Partovi}, which
has numerous advantages over previous formulations. The {\it universal uncertainty relation} (UUR)
based on majorization was found in \cite{Friedland, Puchala}
\begin{equation}\label{e:UUR}
p(\rho)\otimes q(\rho)\prec \omega, \quad  \forall ~ \rho,
\end{equation}
where $\rho$ is any mixed state on a finite dimensional Hilbert space and $\omega$
is a certain probability vector independent of $\rho$.
Here the {\it majorization} '$\prec$ ' is certain partial order among real vectors \cite{Marjorization}
(see section 2 for definition). In the following we will also generalize the UUR principle to
other situations and use them to give new entanglement tests for $k$-separability.

The paper is roughly organized as follows. First we review some basic backgrounds of
majorization theory and the UUR in Sect. 2. Then we prove a universal
lower bound for partial separable states and use the lower bound to give criteria
for genuine entanglement and $k$-separability. Finally we discuss the matrix form of majorization
and how it is used to provide better tests for entanglement of multipartite mixed states.

\section{Background materials} 

We review some basic materials of majorization theory \cite{Marjorization}, the UUR,
and the elementary classification of mixed three-qubit states following mostly \cite{Acin}.

A real vector $x\in\mathds{R}^{d}$ is {\it majorized by} (denoted as $\prec$)
another real vector ${y\in\mathds{R}^{d}}$ provided that $\sum_{j=1}^{k}x_{j}^{\downarrow}\leqslant \sum_{j=1}^{k}y_{j}^{\downarrow}$ for all $1\leqslant k \leqslant d-1$ and $\sum_{j=1}^{d}x_{j}^{\downarrow}= \sum_{j=1}^{d}y_{j}^{\downarrow}$. The down-arrow means to rearrange the components of the vector
in the decreasing order: $x_{1}^{\downarrow}\geqslant x_{2}^{\downarrow}\geqslant \cdots \geqslant x_{d}^{\downarrow}$. If $x\prec y$ but $x\neq y$, then $x$ is said to be strictly majorized by $y$ and written as $x\prec\prec y$.
In general, any probability vectors $x\in\mathds{R}^{d}$ satisfies that
$(\frac{1}{d}, \frac{1}{d}, \cdots, \frac{1}{d})\prec x \prec (1, 0, \cdots, 0)$.

Consider an $n$-partite density matrix $\rho$ ($n\geqslant2$) with positive operator valued measures (POVMs).
Let $\{X_l, E_{\alpha_{l}}^{X_l}\}_{\alpha_{l}=1}^{N_{l}}$ be the $l$-th POVM, where $1\leqslant l\leqslant n$
and $N_{l}$ is the number of elements of the $l$-th POVM. A measurement of $\rho$ with the $l$-th POVM $X_{l}$ induces a probability distribution vector $p^{l}(\rho)=(p^{l}_{1}(\rho),p^{l}_{2}(\rho),\cdots,p^{l}_{N_{l}}(\rho))$,
where $p^l_j(\rho)=tr(\rho E_{\alpha_{j}}^{X_l})$.
Then a uncertainty of the form
\begin{equation}\label{e:MUUR}
\bigotimes_{l=1}^{n}p^{l}(\rho)\prec {\bf \omega}, \quad \forall ~ \rho,
\end{equation}
holds,
where the LHS represents the joint probability distribution induced by measuring $\rho$ with each POVM $X_{l}$.
The multitensor product is defined by associativity as follows. Suppose $a\in \mathbb R^m$
and $b\in\mathbb R^n$, then the tensor product $a\otimes b$
is the vector $(a_1b_1, \cdots, a_1b_n, \cdots, a_mb_n)$ in $\mathbb R^{mn}$.

The vector $\omega$ is independent of $\rho$. If the measurement
elements $X_{l}$ do not have a common eigenstate, then $\omega\prec\prec (1, 0, \cdots, 0)$.
Moreover for any uncertainty
measure $\Phi$, a nonnegative Schur-concave function, one has that
\be\label{phi}
\Phi[\bigotimes_{l=1}^{n}p^{l}(\rho)]\geqslant \Phi(\omega), \quad \forall ~ \rho.
\ee

For example, let $\rho$ be any bipartite density matrix and  ${|a_m\rangle}$,
${|b_m\rangle}$ $(m=1, \ldots, d)$ be two orthonormal bases of the underlying Hilbert space.
Let $p_m(\rho)=\langle a_m|\rho|a_m\rangle$ and $q_n(\rho)=\langle b_n|\rho|b_n\rangle$ be the measurements of $\rho$ given by the basis elements, then they
form two probability distribution vectors ${\bf p}(\rho)$ and ${\bf q}(\rho)$ respectively.
It can be shown that the realignment vector of the Kronecker tensor product of
the two probability vectors ${\bf p}(\rho)$ and ${\bf q}(\rho)$
is majored by a vector ${\bf \omega}$ independent of $\rho$:
\begin{equation}\label{e:PQUUR}
{\bf p}(\rho)\otimes {\bf q}(\rho)\prec {\bf \omega}, \quad \forall ~ \rho,
\end{equation}
where ${\bf \omega}\in \mathbb R^{d^2}$ is given by
\begin{equation}\label{e:omega1}
\omega = (\Omega_1, \Omega_2-\Omega_1, \ldots, \Omega_d-\Omega_{d-1}, 0, \ldots, 0),
\end{equation}
and
\begin{equation}\label{e:omega2}
\Omega_k=\max\limits_{I_k}\max\limits_{\rho}\sum\limits_{(m,n)\in I_{k}}p_m(\rho)\,q_n(\rho).
\end{equation}
Here $I_k$ are $k$-element subsets of $[d]\times [d]$ and $[d]=\{1, 2, \ldots, d\}$.
The outer maximum is over all subsets $I_k$ and the inner maximum runs over all density matrices.

Mixed states of three-qubit systems can be classified by the following: (1) the convex hull $S$ of separable states; (2) the convex hull $B$ of biseparable states (AB-C, AC-B and BC-A); (3) the convex hull of all states, including $S$, $B$ and genuinely entangled states (W-states and GHZ-states). All these sets are compact and convex, satisfy $S\subset B\subset W\subset GHZ$ (see FIG 1.). Examples of GHZ witness and W witness had been given in \cite{Acin}.

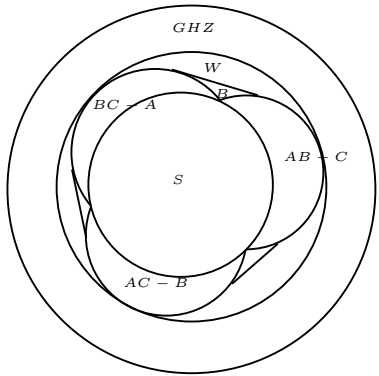
\begin{figure}
\begin{picture}(0,0)%
\includegraphics{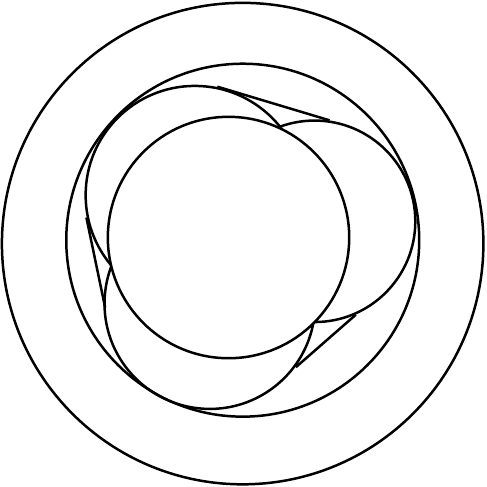}%
\end{picture}%
\setlength{\unitlength}{1575sp}%
\begingroup\makeatletter\ifx\SetFigFont\undefined%
\gdef\SetFigFont#1#2#3#4#5{%
  \reset@font\fontsize{#1}{#2pt}%
  \fontfamily{#3}\fontseries{#4}\fontshape{#5}%
  \selectfont}%
\fi\endgroup%
\begin{picture}(5838,5836)(2967,-7027)
\put(4816,-5641){\makebox(0,0)[lb]{\smash{{\SetFigFont{5}{6.0}{\rmdefault}{\mddefault}{\updefault}{\color[rgb]{0,0,0}$AC-B$}%
}}}}
\put(7336,-3661){\makebox(0,0)[lb]{\smash{{\SetFigFont{5}{6.0}{\rmdefault}{\mddefault}{\updefault}{\color[rgb]{0,0,0}$AB-C$}%
}}}}
\put(4321,-2851){\makebox(0,0)[lb]{\smash{{\SetFigFont{5}{6.0}{\rmdefault}{\mddefault}{\updefault}{\color[rgb]{0,0,0}$BC-A$}%
}}}}
\put(5581,-1636){\makebox(0,0)[lb]{\smash{{\SetFigFont{5}{6.0}{\rmdefault}{\mddefault}{\updefault}{\color[rgb]{0,0,0}$GHZ$}%
}}}}
\put(6256,-2671){\makebox(0,0)[lb]{\smash{{\SetFigFont{5}{6.0}{\rmdefault}{\mddefault}{\updefault}{\color[rgb]{0,0,0}$B$}%
}}}}
\put(6076,-2266){\makebox(0,0)[lb]{\smash{{\SetFigFont{5}{6.0}{\rmdefault}{\mddefault}{\updefault}{\color[rgb]{0,0,0}$W$}%
}}}}
\put(5581,-4021){\makebox(0,0)[lb]{\smash{{\SetFigFont{5}{6.0}{\rmdefault}{\mddefault}{\updefault}{\color[rgb]{0,0,0}$S$}%
}}}}
\end{picture}%
\caption{Schematic picture of mixed states for three qubits. S: Fully separable class; B: Biseparable class (convex hull of biseparable states under three different partitions); W and GHZ: genuine tripartite entangled.}
\end{figure}

\section{ENTANGLEMENT DETECTION}

As shown in \cite{Partovi}, the universal uncertainty relation can reach its bound on pure states. One can use this to establish an entanglement detector to check
whether a density matrix is separable by proving a condition satisfied by all separable states.
The method of using universal uncertainty relations, however, has a limit to detect genuine entanglement. In this section, we utilize the Lagrange multiplier to show that the universal uncertainty bound of $\rho=\sum p_{i} \rho_{AB}^{i}\otimes \rho_{C}^{i}$ can also be reached by the pure state of the form $\rho_{AB}\otimes \rho_{C}$, which provides a AB-C biseparable detector. This establishes a biseparable detector of all
tripartite states
 in AB-C, AC-B or BC-A, and more generally, it can be generalized to
 give a $k$-separable detector for all $n$-partite states.

\subsection{Universal Uncertainty Bounds on Pure States} 

For simplicity we limit ourselves to three measurements $(X,\{E_{\alpha}^{X}\})$, $(Y,\{E_{\beta}^{Y}\})$, and $(Z,\{E_{\gamma}^{Z}\})$ that have no common eigenstate. The measurements are performed on a given state $\rho\in \mathds{H}_{A}\otimes \mathds{H}_{B}\otimes \mathds{H}_{C}$. Let $\mathcal{P}^{X}(\rho)=(tr(\rho E_{\alpha}^{X})), \mathcal{P}^{Y}(\rho)=(tr(\rho E_{\beta}^{Y})), \mathcal{P}^{Z}(\rho)=(tr(\rho E_{\gamma}^{Z}))$ and  $\mathcal{P}^{X\oplus Y\oplus Z}(\rho)=\mathcal{P}^{X}(\rho)\otimes \mathcal{P}^{Y}(\rho)\otimes \mathcal{P}^{Z}(\rho)$.

We recall the uncertainty bound given in \cite{Hossein}. The tensor product of the probability distribution vectors satisfies that
\begin{equation}
\mathcal{P}^{X}(\rho)\otimes \mathcal{P}^{Y}(\rho)\otimes \mathcal{P}^{Z}(\rho)\prec \omega,
\end{equation}
where the bound is reached on pure states $\rho_{0}$:
\begin{equation}
\omega=\sup\limits_{ \rho_{0}}[\mathcal{P}^{X\oplus Y\oplus Z} \rho_{0})]\prec\prec (1,0, \cdots, 0).
\end{equation}
If $\rho=\sum p_{i} \rho_{A}^{i}\otimes \rho_{B}^{i}\otimes \rho_{C}^{i}$, then
\begin{align}
\mathcal{P}^{X}(\rho)\otimes \mathcal{P}^{Y}(\rho)\otimes \mathcal{P}^{Z}(\rho)
\prec \omega_{A, B,C}=\sup\limits_{\rho_{A}\otimes\rho_{B}\otimes\rho_{C}}[\mathcal{P}^{X\oplus Y\oplus Z}(\rho_{A}\otimes\rho_{B}\otimes\rho_{C})],
\end{align}
where $\rho_{A}\otimes\rho_{B}\otimes\rho_{C}$ are fully separable pure states.
Now we want to prove that if  the density matrix of a tripartite states has the form of $\rho=\sum p_{i} \rho_{AB}^{i}\otimes \rho_{C}^{i}$, then
\begin{align}
\mathcal{P}^{X}(\rho)\otimes \mathcal{P}^{Y}(\rho)\otimes \mathcal{P}^{Z}(\rho)
&\prec \omega_{AB, C}=\sup\limits_{\rho_{AB}\otimes\rho_{C}}[\mathcal{P}^{X\oplus Y\oplus Z}(\rho_{AB}\otimes\rho_{C})],
\end{align}
where $\rho_{AB}\otimes\rho_{C}$ is biseparable pure state of type AB-C. To prove this, it is sufficient to show that the maximum value of the sum of $i$ different components of $\mathcal{P}^{X\oplus Y\oplus Z}(\sum p_{i} \rho_{AB}^{i}\otimes \rho_{C}^{i})$ is realized on pure states of same type.

Suppose $\rho=\sum\limits_{a}q_{a}|\phi_{a}^{AB}\rangle\langle\phi_{a}^{AB}|\otimes|\phi_{a}^{C}\rangle\langle\phi_{a}^{C}|$.
To maximize $\mathcal{P}^{X\oplus Y\oplus Z}(\rho)$ we apply the method of the Lagrange multipliers:
\begin{widetext}
\begin{equation}
\begin{aligned}
\sum\limits_{a,b,c}q_{a}q_{b}q_{c}[\langle\phi_{a}^{AB}|\langle\phi_{a}^{C}|E_{\alpha_{1}}^{X}|\phi_{a}^{C}\rangle|\phi_{a}^{AB}\rangle
\langle\phi_{b}^{AB}|\langle\phi_{b}^{C}|E_{\beta_{1}}^{Y}|\phi_{b}^{C}\rangle|\phi_{b}^{AB}\rangle
\langle\phi_{c}^{AB}|\langle\phi_{c}^{C}|E_{\gamma_{1}}^{Z}|\phi_{c}^{C}\rangle|\phi_{c}^{AB}\rangle+\\
\langle\phi_{a}^{AB}|\langle\phi_{a}^{C}|E_{\alpha_{2}}^{X}|\phi_{a}^{C}\rangle|\phi_{a}^{AB}\rangle
\langle\phi_{b}^{AB}|\langle\phi_{b}^{C}|E_{\beta_{2}}^{Y}|\phi_{b}^{C}\rangle|\phi_{b}^{AB}\rangle
\langle\phi_{c}^{AB}|\langle\phi_{c}^{C}|E_{\gamma_{2}}^{Z}|\phi_{c}^{C}\rangle|\phi_{c}^{AB}\rangle+ \cdots +\\
\langle\phi_{a}^{AB}|\langle\phi_{a}^{C}|E_{\alpha_{i}}^{X}|\phi_{a}^{C}\rangle|\phi_{a}^{AB}\rangle
\langle\phi_{b}^{AB}|\langle\phi_{b}^{C}|E_{\beta_{i}}^{Y}|\phi_{b}^{C}\rangle|\phi_{b}^{AB}\rangle
\langle\phi_{c}^{AB}|\langle\phi_{c}^{C}|E_{\gamma_{i}}^{Z}|\phi_{c}^{C}\rangle|\phi_{c}^{AB}\rangle]+\\
\sum\limits_{a}\eta_{i,a}^{AB}(1-\langle\phi_{a}^{AB}|\phi_{a}^{AB}\rangle)+
\sum\limits_{a}\eta_{i,a}^{C}(1-\langle\phi_{a}^{C}|\phi_{a}^{C}\rangle)+
\xi_{i}(1-\sum\limits_{a}q_{a}).
\end{aligned}
\end{equation}
\end{widetext}
where $\eta_{i, a}^{AB}, \eta_{i,a}^{C}, \xi_{i}$ are the Lagrange multipliers.

Denote that
\begin{align}
\varepsilon_{i}^{X}&=\sum\limits_{k=1}^{i}\mathcal{P}_{\beta_{k}}^{Y}(\rho)\mathcal{P}_{\gamma_{k}}^{Z}(\rho)E_{\alpha_{k}}^{X}\notag,\\
\varepsilon_{i}^{Y}&=\sum\limits_{k=1}^{i}\mathcal{P}_{\alpha_{k}}^{X}(\rho)\mathcal{P}_{\gamma_{k}}^{Z}(\rho)E_{\beta_{k}}^{Y}\notag,\\
\varepsilon_{i}^{Z}&=\sum\limits_{k=1}^{i}\mathcal{P}_{\alpha_{k}}^{X}(\rho)\mathcal{P}_{\beta_{k}}^{Y}(\rho)E_{\gamma_{k}}^{Z}\notag,\\
\varepsilon_{i}&=\varepsilon_{i}^{X}+\varepsilon_{i}^{Y}+\varepsilon_{i}^{Z}.
\end{align}

Variations with respect to $\langle\phi_{s}^{AB}|$ give
\begin{equation}
q_{s}[\langle\phi_{s}^{c}|\varepsilon_{i}|\phi_{s}^{C}]|\phi_{s}^{AB}\rangle=\eta_{i,s}^{AB}|\phi_{s}^{AB}\rangle.
\end{equation}
Similarly with respect to $\langle\phi_{s}^{C}|$ we have that
\begin{equation}
q_{s}[\langle\phi_{s}^{AB}|\varepsilon_{i}|\phi_{s}^{AB}]|\phi_{s}^{C}\rangle=\eta_{i,s}^{C}|\phi_{s}^{C}\rangle,
\end{equation}
which implies $\eta_{i,s}^{AB}=\eta_{i,s}^{C}$. Denote $\Omega_{AB,C;i}$ as the sum of $i$ different components, and $\langle\phi_{s}^{AB}|\langle\phi_{s}^{C}|\varepsilon_{i}|\phi_{s}^{C}\rangle|\phi_{s}^{AB}\rangle=\varepsilon_{i,ss}$, then
$\Omega_{AB,C;i}=\frac{1}{3}\sum\limits_{s}q_{s}\varepsilon_{i,ss}=\frac{1}{3}tr(\varepsilon_{i}\rho)$. Finally, variation with respect to $q_{s}$ leads to
$\langle\phi_{s}^{AB}|\langle\phi_{s}^{C}|\varepsilon_{i}|\phi_{s}^{C}\rangle|\phi_{s}^{AB}\rangle=\xi_{i}$ which means that
$\varepsilon_{i,ss}$ do not depend on $s$, and $\omega_{AB,C}$ can be reached by pure biseparable states of form $\rho_{AB}\otimes\rho_{C}$.

It is clear that the above argument works for any multipartite state, and the bound for the tensor product of
multipartite state can be reached by pure states with the same type. This implies that majorization uncertainty relations can be used to detect genuine entanglement in multipartite and distinguish different types
at the same level of entanglement (for example, triseparable ABC-D vs. quartistate system).

\subsection{Entanglement Detection in Tripartite States} 

Let $\rho\in \mathds{H}_{A}\otimes\mathds{H}_{B}\otimes\mathds{H}_{C}$ be a tripartite state
with a measurement $\{X,E_{\alpha}^{X}\}$. Then the probability vector $\mathcal{P}^{X}(\rho)$ is majorized by a bound vector $\omega$ independent of $\rho$. The following results are clear from our discussion.

\noindent [{\sf Lemma 1}].
Let $\rho=\sum\limits_{i}p_{i}\rho_{A}^{i}\otimes\rho_{B}^{i}\otimes\rho_{C}^{i}$ be a mixed tripartite state. The probability density
vector $\mathcal{P}^{X}(\rho)$ associated with the measurement $X$ is majorized by a bound vector $\omega_{A, B, C}$ independent from $\rho$
and reachable by a pure tripartite state:
\begin{equation}
\mathcal{P}^{X}(\rho)\prec\omega_{A, B, C}.
\end{equation}
If the probability vector of $\rho$ violates the majorization relation, then $\rho$ is entangled.

\noindent [{\sf Lemma 2}]. If $\rho=\sum\limits_{i}p_{i}\rho_{AB}^{i}\otimes\rho_{C}^{i}$, then the probability vector $\mathcal{P}^{X}(\rho)$ resulting from the measurement $X$ on $\rho$ is majorized by $\omega_{AB, C}$ which is state-independent and reachable by some pure biseparable state of type AB-C:
\begin{equation}
\mathcal{P}^{X}(\rho)\prec\omega_{AB, C}.
\end{equation}
If the probability vector of $\rho$ violates the majorization relation, then
$\rho$ is entangled but not biseparable of type AB-C.

If the dimensions of $\mathds{H}_{A}$, $\mathds{H}_{B}$ and $\mathds{H}_{C}$ are different, then the majorization uncertainty bounds $\omega_{AB, C}$, $\omega_{AC, B}$ and $\omega_{BC, A}$ are all different under a suitable measurement $\{X,E_{\alpha}^{X}\}$. These probability vectors and
that of the whole space form a lattice (see FIG 2), which leads to a majorization uncertainty bound
to control any two of $\omega_{AB, C}$, $\omega_{AC, B}$ and $\omega_{BC, A}$.

\begin{figure}
\begin{picture}(0,0)%
\includegraphics{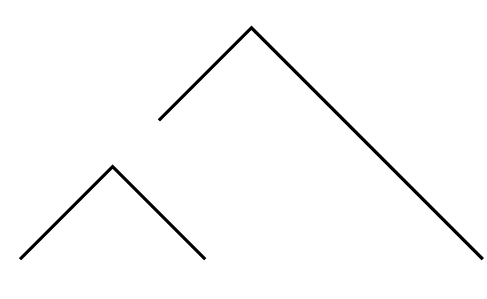}%
\end{picture}%
\setlength{\unitlength}{1948sp}%
\begingroup\makeatletter\ifx\SetFigFont\undefined%
\gdef\SetFigFont#1#2#3#4#5{%
  \reset@font\fontsize{#1}{#2pt}%
  \fontfamily{#3}\fontseries{#4}\fontshape{#5}%
  \selectfont}%
\fi\endgroup%
\begin{picture}(4728,2883)(1606,-7645)
\put(3871,-4921){\makebox(0,0)[lb]{\smash{{\SetFigFont{6}{7.2}{\rmdefault}{\mddefault}{\updefault}{\color[rgb]{0,0,0}$\omega_{123}$}%
}}}}
\put(2521,-6226){\makebox(0,0)[lb]{\smash{{\SetFigFont{6}{7.2}{\rmdefault}{\mddefault}{\updefault}{\color[rgb]{0,0,0}$\omega_{23}$}%
}}}}
\put(1621,-7576){\makebox(0,0)[lb]{\smash{{\SetFigFont{6}{7.2}{\rmdefault}{\mddefault}{\updefault}{\color[rgb]{0,0,0}$\omega_{AB, C}$}%
}}}}
\put(3421,-7576){\makebox(0,0)[lb]{\smash{{\SetFigFont{6}{7.2}{\rmdefault}{\mddefault}{\updefault}{\color[rgb]{0,0,0}$\omega_{AC, B}$}%
}}}}
\put(6166,-7576){\makebox(0,0)[lb]{\smash{{\SetFigFont{6}{7.2}{\rmdefault}{\mddefault}{\updefault}{\color[rgb]{0,0,0}$\omega_{BC, A}$}%
}}}}
\end{picture}%
\caption{Preordering of $\omega_{AB, C}$, $\omega_{AC, B}$ and $\omega_{BC, A}$.}
\end{figure}
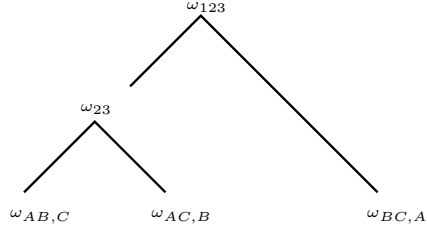

\noindent [{\sf Proposition 1}]. If $\omega_{AB, C}\neq\omega_{AC, B}$ and both $\prec\prec (1, 0, \cdots, 0)$,
 then there exists a unique probability vector $\omega_{23}$ such that
\begin{align}\label{e:UUR23}
\omega_{AB, C}&\prec\omega_{23},\notag\\
\omega_{AC, B}&\prec\omega_{23},\\
\omega_{23}\prec\prec (1&, 0, \cdots, 0), \notag
\end{align}
and majorizied by any other vector with property Eq. (\ref{e:UUR23}). This vector is also denoted by
$\max\{\omega_{AB, C},\omega_{AC, B}\}$.

With these results in hand, we can give an entanglement criterion to analyze biseparable states.

\noindent [{\sf Theorem 1}]. If the probability vector of tripartite state $\rho$ under measurement $\{X,E_{\alpha}^{X}\}$ is
not majorized by 
$\omega_{23}$:
\begin{equation}
\mathcal{P}^{X}(\rho)\nprec\max\{\omega_{AB, C},\omega_{AC, B}\}, 
\end{equation}
then $\rho$ can not be of the form $\rho=\sum\limits_{i}p_{i}\rho_{AB}^{i}\otimes\rho_{C}^{i}+\sum\limits_{j}q_{j}\rho_{AC}^{i}\otimes\rho_{B}^{i}$.

To study genuine entanglement, we consider both $\omega_{23}$
and $\omega_{BC, A}$
and denote the unique probability vector $\max\{\omega_{23}, \omega_{BC, A}\}$ by $\omega_{123}$.
The following result gives a criterion
for genuine entanglement.

\noindent [{\sf Theorem 2}]. If
\begin{equation}
\mathcal{P}^{X}(\rho)\nprec\omega_{123},
\end{equation}
then $\rho$ can not be biseparable, and $\rho$ is genuinely entangled.

For any measurement $\Phi$ of uncertainty (non-negative Schur-concave function), it follows from $\mathcal{P}^{X}(\rho)\prec\omega_{123}$ that
\begin{equation}\label{e:NUUR123}
\Phi(\mathcal{P}^{X}(\rho))\geqslant \Phi(\omega_{123}).
\end{equation}

One can use uncertainty measurements to give numerical criteria for genuine entanglement.

\noindent [{\sf Corollary 1}]. If there is an uncertainty measurement $\Phi$ such that
\begin{equation}
\Phi(\mathcal{P}^{X}(\rho))< \Phi(\omega_{123}),
\end{equation}
then $\rho$ is not a biseparable state, and $\rho$ is genuinely entangled.

If $\mathds{H}_{A}$, $\mathds{H}_{B}$ and $\mathds{H}_{C}$ have the same dimension, then
$\omega_{AB, C}=\omega_{AC, B}=\omega_{BC, A}=\omega_{123}$
by the symmetry property of the superior over pure biseparable states. Although one can not use
the above result to detect if a tripartite state has the form $\rho=\sum\limits_{i}p_{i}\rho_{AB}^{i}\otimes\rho_{C}^{i}+\sum\limits_{j}q_{j}\rho_{AC}^{i}\otimes\rho_{B}^{i}$,
one still has the following result for genuine entanglement.

\noindent [{\sf Theorem 3}]. For $\dim\mathds{H}_{A}=\dim\mathds{H}_{B}=\dim\mathds{H}_{C}$, if
\begin{equation}
\mathcal{P}^{X}(\rho)\nprec\omega_{AB, C},
\end{equation}
then $\rho$ is not a biseparable state, and $\rho$ is genuinely entangled.

Combining with non-negative Schur-concave functions, one immediately gets
the following criterion of
genuine entanglement.

\noindent [{\sf Corollary 2}]. If there is a non-negative Schur-concave function $\Phi$ such that
\begin{equation}
\Phi(\mathcal{P}^{X}(\rho))< \Phi(\omega_{AB, C}),
\end{equation}
then $\rho$ is not biseparable state, and the state $\rho$ is genuinely entangled.

\subsection{Entanglement Detection in Multipartite States} 

The majorization method can be generalized to characterize entanglement for multipartite quantum states. In this
section, we present a general and systematic scheme to study $k$-separable and  $n$-partite genuinely entangled
states. 

Following \cite{Huber}, we say a pure $n$-partite quantum state $|\Psi\rangle$ is $k$-separable if it can be written as
\begin{equation}
|\Psi\rangle=|\phi_{1}\rangle\otimes|\phi_{2}\rangle\otimes\cdots\otimes|\phi_{k}\rangle,
\end{equation}
where $|\phi_{i}\rangle$ is a single subsystem or a group of subsystems. If such
as decomposition is not possible, 
 $|\Psi\rangle$ is called genuinely $n$-partite entangled. For a mixed state $\rho$: a state is genuinely $k$-partite entangled if any decomposition into pure states
\begin{equation}
\rho=\sum\limits_{i}p_{i}|\psi_{i}\rangle\langle\psi_{i}|,
\end{equation}
with probabilities $p_{i}>0$ contains at least one genuinely $k$-partite entangled component. It
is called $k$-separable if any decomposition into pure states $|\psi_{i}\rangle$ is at least
$k$-separable. All $(k+1)$-separable states form a compact and convex subset of the set of $k$-separable states.
This hierarchical structure is usually referred to as a {\it uniform entanglement frame}.
Our main idea is to construct criteria to detect $k$-separability by majorization uncertainty relations and
find the principle behind these criteria.

Statistical mixture of biseparable states for tripartite states has been considered,
we now focus on $k$-separable states for multipartite states.
For any $k$-separable state $\rho_{kse}=\sum\limits_{i}p_{i}\rho_{kse}^{i}$, if all
$\rho_{kse}^{i}$ have the same form (for example, systems $1$ to $n-k+1$ are entangled and the rest are separable), then by the majorization method there exists a bound vector $\omega_{k,l}$
reachable by pure states 
with the same form as $\rho_{kse}^{i}$ for each $i$. Moreover,
\begin{equation}
\mathcal{P}^{X}(\rho_{kse})\prec\omega_{k,l},
\end{equation}
where there are $\mathbf{C}_{n}^{k}$ possibilities for $l$. If $\dim\mathds{H}_{1}=\dim\mathds{H}_{2}=\cdots=\dim\mathds{H}_{n}$, then $$\omega_{k,1}=\omega_{k,2}=\cdots=\omega_{k,\mathbf{C}_{n}^{k}}:=\omega_{k}.$$
Under this assumption we have
\medskip

\noindent [{\sf Theorem 4}]. For any $k$-separable state $\rho_{kse}$ and
measurement $\{X,E_{\alpha_{i}}^{X}\}$, the following majorization uncertainty relation holds:
\begin{equation}
\mathcal{P}^{X}(\rho_{kse})\prec\omega_{k},
\end{equation}
where $\omega_{k}$ can be reached by pure $k$-separable state.
\medskip

\noindent [{\sf Corollary 3}]. For any $n$-partite state $\rho$, if there
exists a measurement $\{X,E_{\alpha_{i}}^{X}\}$ such that the probability vector is not majorized by $\omega_{k}$:
\begin{equation}
\mathcal{P}^{X}(\rho)\nprec\omega_{k},
\end{equation}
then $\rho$ is at most $(k-1)$-separable.
\medskip

One can generalize Corollary 3 to derive a criterion
for genuinely $n$-partite entangled states.

\noindent [{\sf Corollary 4}]. For any $n$-partite state $\rho$, if there exists
a probability vector under measurement $\{X,E_{\alpha_{i}}^{X}\}$ such that
\begin{equation}\label{e:UUR2}
\mathcal{P}^{X}(\rho)\nprec\omega_{2},
\end{equation}
then $\rho$ is genuinely $n$-partite entangled.

Under non-negative Schur-concave functions $\Phi$, the majorization uncertainty bound $\omega_{2}$ becomes a real number $\Phi(\omega_{2})$, then Eq. (\ref{e:UUR2}) generates in fact infinite family of genuinely $n$-partite entangled criteria.

\noindent [{\sf Corollary 5}]. For any $n$-partite state $\rho$ and non-negative Schur-concave functions $\Phi$, if the following equality holds under measurement $\{X,E_{\alpha_{i}}^{X}\}$:
\begin{equation}
\Phi(\mathcal{P}^{X}(\rho))<\Phi(\omega_{2}),
\end{equation}
then $\rho$ is genuinely $n$-partite entangled.
\medskip

Entangled states are characterized well under the majorization uncertainty relations as
the universal uncertainty bounds are independent from the state $\rho$. This is particularly so
when one applies the uncertainty relations given in Eq. (\ref{e:H1}). 
We discuss some examples to show how these are applied.

\noindent [{\sf Example 1}]. Consider the Werner state $\rho^{wer}_{d}(q)$ \cite{Hossein} defined on
the tensor product of two $d$-dimensional Hilbert spaces:
\begin{equation}
\rho^{wer}_{d}(q)=\frac{1}{d^2}(1-q)I+q|\mathfrak{B}_{1}\rangle\langle\mathfrak{B}_{1}|,
\end{equation}
where $I$ is the identity matrix in $\mathbb C^{d\times d}$ and
\begin{equation}
|\mathfrak{B}_{1}\rangle=\frac{1}{\sqrt{d}}\sum\limits_{j=0}^{d-1}|j_{A}\rangle\otimes|j_{B}\rangle,
\end{equation}
is the first of the generalized Bell states $\{|\mathfrak{B}_{\alpha}\rangle\}_{\alpha=1}^{d^{2}}$,
which form a basis of orthonormal eigenstates for $\rho^{wer}_{d}(q)$.
Let the measurement $\{X, E_{\alpha}^{X}\}$ be $E_{\alpha}^{X}=|\mathfrak{B}_{\alpha}\rangle\langle\mathfrak{B}_{\alpha}|$ , $\alpha=1, 2, \cdots, d^{2}$.
The probability vector is then
\begin{equation}
\mathcal{P}^{X}(\rho^{wer}_{d}(q))=(q+d^{-2}(1-q), d^{-2}(1-q), d^{-2}(1-q), \cdots, d^{-2}(1-q)).
\end{equation}
As shown in \cite{Bourennane}, the maximum overlap of every generalized Bell states with the set of pure, product states is given by $1/\sqrt{d}$, we get that
\begin{equation}
\omega_{A, B}=(1/d, 1/d, \cdots, 1/d, 0, 0, \cdots, 0).
\end{equation}
Using our Theorem 1, the Werner state is separable if $\mathcal{P}^{X}(\rho^{wer}_{d}(q))$ is majorized by $\omega_{A, B}$, i.e., $q\leqslant (1+d)^{-1}$. This inequality agrees
with the well-known separability condition for the Werner state. 

Clearly, the efficiency of uncertainty bounds in judging entanglement depends on the choice of the measurement $\{X, E_{\alpha}^{X}\}$. The calculation of $\omega$ will be easy if the measurement is optimal in some sense.

\noindent [{\sf Example 2}]. Next we will
take the quantum state to be $\rho^{wer}_{d}(q)\otimes\frac{1}{d}I$ and the measurement to be $\{X, E_{\alpha, j}^{X}\}$, where $E_{\alpha, j}^{X}=|\mathfrak{B}_{\alpha}\rangle\otimes|j\rangle$, $\alpha=1, 2, \cdots, d^{2}$; $j=0, 1, \cdots, d-1.$ It is easy to see that
\begin{align}
&\mathcal{P}^{X}(\rho^{wer}_{d}(q)\otimes\frac{1}{d}I)\notag\\
=&(\underbrace{q+d^{-2}(1-q), \cdots, q+d^{-2}(1-q)}_{d-times}, d^{-2}(1-q), d^{-2}(1-q), \cdots, d^{-2}(1-q))\notag.
\end{align}
The maximum overlap of every measurement element $E_{\alpha, j}^{X}=|\mathfrak{B}_{\alpha}\rangle\otimes|j\rangle$ with the set of pure, biseparable state is given by $1$. Thus we conclude that
\begin{equation}
\omega_{AB, C}=(1, 0, \cdots, 0).
\end{equation}
It is easy to see that $\mathcal{P}^{X}(\rho^{wer}_{d}(q)\otimes\frac{1}{d}I)\prec\omega_{AB, C}$ is always ture, and
no violation happens, which means that $\rho^{wer}_{d}(q)\otimes\frac{1}{d}I$ is always biseparable of form AB-C. This coincides with the fact that for any value $q$, $\rho^{wer}_{d}(q)\otimes\frac{1}{d}I$ is always biseparable and
can only be entangled in the AB system.

\section{Matrix Forms of Majorization} 

Majorization was first studied by Schur \cite{Schur} in relation with
Hadamard inequalities and it was proved that two probability vectors
$x\prec y$ if and only if  $y=Q(x)$ for a bistochastic matrix $Q$.
A matrix $Q=(q_{ij})\in\mathbb R_+^{n\times n}$ is bistochastic if
$$\sum_iq_{ij}=\sum_jq_{ij}=1.$$
Birkkhoff proved that the set of all bistochastic matrices is the convex hull of
permutation matrices \cite{Ando}.
The geometric and combinatoric principle behind majorization and bistochastic matrix is that the relation fully characterizes the convexity of the set of all $k$-separable states. It is a convex and compact subset of $(k-1)$-separable states, which
enables one to construct criteria to detect $k$-separable states.
The majorization method to detect states is relied on the ``interior'' of
the convex set of $k$-separable states based on Brikhoff's theorem.

We list the equivalent forms of majorization as follows.

(1) $x\prec y$;

(2) $y=Q(x)$ for some bistochastic matrix $Q$;  

(3) $\Phi(x)\geqslant \Phi(y)$ for any non-negative Schur-concave functions $\Phi$;

(4) $x$ can be derived from $y$ by successive applications of finitely many $T$-transformations:
\begin{equation}
T=\lambda I+(1-\lambda)P,
\end{equation}
where $0\leqslant\lambda\leqslant1$ and $P$ is a permutation matrix that interchanges two coordinates;

(5) $x$ is in the convex hull of the $n!$ permutations of $y$.

We now give the matrix form of the $k$-separable criterion.

\noindent [{\sf Theorem 5}]. Let $\rho$ be an $n$-partite quantum state
on $\mathds{H}^{\otimes n}$ ($\dim(\mathds{H})=d$),
and let $\{X, E_{\alpha}^{X}\}$ be a measurement with the probability
vector $\mathcal{P}^{X}(\rho)=(p_{1}, p_{2}, \cdots, p_{m})$, where $m=d^n$. Suppose the $k$-separable
 uncertainty bound is $\omega_{k}=(\Omega_1, \Omega_2-\Omega_1, \ldots, \Omega_m-\Omega_{m-1})$.
If there does not exist a bistochastic matrix $Q=(q_{ij})$ such that
\begin{align}
\Omega_{j}-\Omega_{j-1}&=\sum\limits_{i=1}^{m}q_{ij}p_{i}\notag, \quad j=1, 2, \cdots, m.
\end{align}
then $\rho$ is at most $(k-1)$-separable.
\medskip

This result follows from the fact that
violation of $\mathcal{P}^{X}(\rho)\prec\omega_{k}$ is equivalent to there does not exist a bistochastic matrix $Q$ such that $\omega_{k}=Q(\mathcal{P}^{X}(\rho))$.

We now consider an information quantity called $f$-relative entropy to express the closeness between two probability vectors and helps describe $k$-separability.  The $f$-relative entropy is also called
$f$-divergence in information theory \cite{Igor}.

\noindent [{\sf Lemma 3}]. Let $f$ be a convex function. Then the information quantity $D_{f}(x\parallel y)=\sum\limits_{i}x_{i}f(\frac{y_{i}}{x_{i}})$ for probability vectors $x=(x_{i})$ and $y=(y_{i})$ is
monotonic: 
\begin{equation}\label{e:frelative}
D_{f}(x\parallel y)\geqslant D_{f}(Q(x)\parallel Q(y)),
\end{equation}
where $Q$ is stochastic matrix.
\medskip

It follows from Lemma 3 that $D_{f}(x\parallel y)$ is an $f$-relative entropy.
This monotonicity condition (\ref{e:frelative}) also holds if $Q$ is replaced by a bistochastic matrix.

Note that $Q(1/m, 1/m, \cdots, 1/m)=(1/m, 1/m, \cdots, 1/m)$ for any bistochastic matrix. Plugging $y=\mathfrak{K}=(1/m, 1/m, \cdots, 1/m)$ and
$x=\mathcal{P}^{X}(\rho)$ associated with a measurement $\{X,E_{\alpha}^{X}\}$ in Eq. (\ref{e:frelative}), one immediately gets that
\begin{equation}
D_{f}(\mathcal{P}^{X}(\rho)\parallel \mathfrak{K})\geqslant D_{f}(\omega_{k}\parallel \mathfrak{K}),
\end{equation}

This gives the following criterion of relevant entanglement.

\noindent [{\sf Theorem 6}]. If there exists a convex function and a measurement $\{X,E_{\alpha}^{X}\}$ such that 
\begin{equation}
D_{f}(\mathcal{P}^{X}(\rho)\parallel \mathfrak{K})< D_{f}(\omega_{k}\parallel \mathfrak{K}),
\end{equation}
then the state $\rho$ is at most $(k-1)$-separable.
\medskip

The $f$-relative entropy $D_{f}(\mathcal{P}^{X}(\rho)\parallel \mathfrak{K})$ expresses the closeness between $\mathcal{P}^{X}(\rho)$ and a fixed point $\mathfrak{K}$, but it does not satisfy
the axioms of a distance. To get a 'distance' to characterize $k$-separability one needs special convex functions such as $f(x)=1-\sqrt{x}$, its square root is called Hellinger distance and denoted by $d_{2}(x, y)$.
It is known that $d_{2}(x, y)$ satisfies the axioms of a distance. With $\mathfrak{K}$ as the center of the circle, $d_{2}(\omega_{k}, \mathfrak{K})$ can be viewed
as the radius of a $k$-separable circle, or simply the $k$-circle ( see FIG 3.).

\begin{figure}
\begin{picture}(0,0)%
\includegraphics{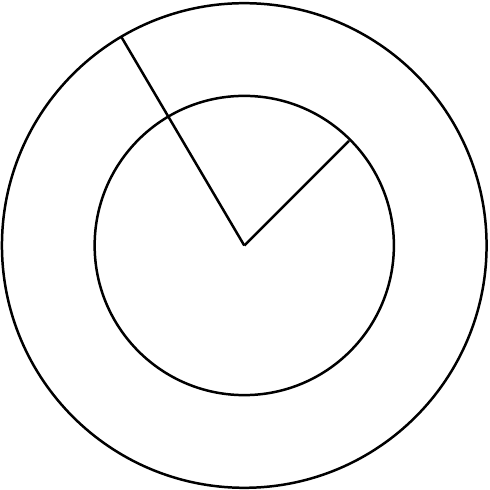}%
\end{picture}%
\setlength{\unitlength}{1575sp}%
\begingroup\makeatletter\ifx\SetFigFont\undefined%
\gdef\SetFigFont#1#2#3#4#5{%
  \reset@font\fontsize{#1}{#2pt}%
  \fontfamily{#3}\fontseries{#4}\fontshape{#5}%
  \selectfont}%
\fi\endgroup%
\begin{picture}(5876,5876)(2913,-7049)
\put(6301,-3841){\makebox(0,0)[lb]{\smash{{\SetFigFont{6}{7.2}{\rmdefault}{\mddefault}{\updefault}{\color[rgb]{0,0,0}$d_{2}(\omega_{k}, \mathfrak{K})$}%
}}}}
\put(5716,-4336){\makebox(0,0)[lb]{\smash{{\SetFigFont{6}{7.2}{\rmdefault}{\mddefault}{\updefault}{\color[rgb]{0,0,0}$\mathfrak{K}$}%
}}}}
\put(4681,-2041){\makebox(0,0)[lb]{\smash{{\SetFigFont{6}{7.2}{\rmdefault}{\mddefault}{\updefault}{\color[rgb]{0,0,0}$d_{2}(\omega_{k+1}, \mathfrak{K})$}%
}}}}
\end{picture}%
\caption{$k$-separable radius corresponds to ' $k$-circle '.}
\end{figure}
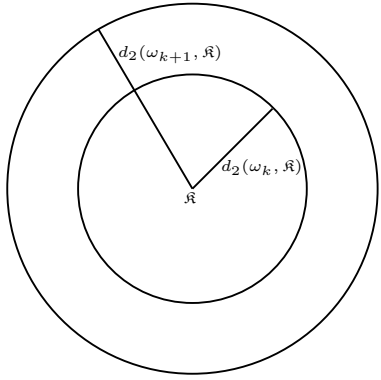

\noindent [{\sf Corollary 5}]. If $\rho$ is at most $(k-1)$-separable, then
\begin{equation}
d_{2}(\mathcal{P}^{X}(\rho), \mathfrak{K})< d_{2}(\omega_{k}, \mathfrak{K}).
\end{equation}
The vector  $\mathcal{P}^{X}(\rho)\in\mathds{R}^{m}$ is a point inside the $k$-separable circle. 

\noindent [{\sf Example 3}]. Consider again the Werner state $\rho^{wer}_{d}(q)$. Both the probability $\mathcal{P}^{X}(\rho^{wer}_{d}(q))$ under measurement $E_{\alpha}^{X}=|\mathfrak{B}_{\alpha}\rangle\langle\mathfrak{B}_{\alpha}|$  and the uncertainty bound $\omega_{A, B}$ are given as follows.
\begin{equation}
\mathcal{P}^{X}(\rho^{wer}_{d}(q))=(q+d^{-2}(1-q), d^{-2}(1-q),
 d^{-2}(1-q), \cdots, d^{-2}(1-q)),
\end{equation}
while
\begin{equation}
\omega_{A, B}=(1/d, 1/d, \cdots, 1/d, 0, 0, \cdots, 0),
\end{equation}
It is easy to calculate that $d_{2}^{2}(\omega_{A, B}, \mathfrak{K})=1-\sqrt{\frac{1}{d}}$. Then
$d_{2}(\mathcal{P}^{X}(\rho^{wer}_{d}(q)), \mathfrak{K})< d_{2}(\omega_{A, B}, \mathfrak{K})$ implies $q>(1+d)^{-1}$,
which coincides with the necessary and sufficient condition for inseparability of the Werner state.

Hellinger distance is useful to detect entanglement based on the monotonicity condition. However, only the monotonicity property is not good enough for entanglement detection, which is
shown by the following example.

\noindent [{\sf Example 4}]. Consider the variational distance given by
\begin{equation}
d_{1}(x, y)=\frac{1}{2}\sum\limits_{i}|x_{i}- y_{i}|.
\end{equation}
It is not an f-relative entropy, but it satisfies the monotonicity property
\begin{equation}
d_{1}(x, y)\geqslant d_{1}(Q(x), Q(y)).
\end{equation}
If we take $y$ as $\mathfrak{K}=(1/m, \cdots, 1/m)$ then $d_{1}(x, y)= d_{1}(Q(x), Q(y))$. Thus $d_{1}$ fails to detect entanglement.

Finally we would like to discuss how entanglement witness is used to detect entanglement. A {\it witness for genuine $k$-partite entanglement} is an observable that has a positive expectation value on states with $(k-1)$-partite entanglement and a negative expectation value on some $k$-partite entangled states. Entanglement witness is a useful tool for analyzing entanglement in experiments \cite{Bourennane}. Usual
entanglement witness does have its limit in dealing with the type of $k$-partite separable state. We now give a universal entanglement witness 
which can detect the type.
We take tripartite states to illustrate the idea. Suppose $\rho\in \mathds{H}_{A}\otimes\mathds{H}_{B}\otimes\mathds{H}_{C}$ and $\dim\mathds{H}_{A}\neq\dim\mathds{H}_{B}\neq\dim\mathds{H}_{C}$.
By choosing a suitable measurement $\{X, E_{\alpha}^{X}\}$, the bounds $\omega_{AB, C}$, $\omega_{AC, B}$ and $\omega_{BC, A}$ are all different. Write $\omega_{AB, C}$ as
$(\Omega_1, \Omega_2-\Omega_1, \ldots, \Omega_m-\Omega_{m-1})$, $m=\dim\mathds{H}_{A}\cdot\dim\mathds{H}_{B}\cdot\dim\mathds{H}_{C}$.

A universal witness operator can be written in the canonical form
\begin{equation}
\mathcal{W}_{k}=\Omega_{k}I-\sum\limits_{i=1}^{k}E_{\alpha_{i}}^{X}, k=1, 2, \cdots, m,
\end{equation}
where $I$ is the identity operator and $E_{\alpha_{i}}^{X}$ are from the measurement $\{X, E_{\alpha}^{X}\}$. Then for any biseparable state $\rho=\sum\limits_{i}p_{i}\rho_{AB}^{i}\otimes\rho_{C}^{i}$ in type AB-C, we have
\begin{equation}
Tr(\mathcal{W}_{k}\rho)\geqslant0, k=1, 2, \cdots, m.
\end{equation}
If there is
\begin{equation}
Tr(\mathcal{W}_{k}\rho)<0,
\end{equation}
for some $k$, then $\rho$ can not be of type AB-C.
We remark that the universal entanglement witness can be used to analyze the complement of the compact convex subset inside another one.

\section{Conclusions and Discussions} 

In this paper, we have developed entanglement criteria based on both majorization and universal uncertainty relations. These criteria have advantage over the scalar detecting algorithms
as they are often stronger and tighter due to specific formulas for the bound $\omega$ of the uncertainty relation. They are tight enough to detect $k$-separable $n$-partite states and their types
by choosing a suitable measurement $\{X, E_{\alpha}^{X}\}$
when the underlying particles have different dimensions.

We have also presented the matrix form of majorization. One feature in this approach is by choosing a suitable monotonic function one can define a distance in the set of entangled states,
where $k$-separable states form concentric circles. In this regard, viable functions are non-negative
Schur-concave functions such as Shannon, R\'{e}nyi and Tsallis entropies \cite{Tsallis, Rudnicki}.
We have also generalized the entanglement witness to
detect the type of a $k$-separable state, and indicate how they can be implemented in experiments.

\medskip
\noindent{\bf Acknowledgments}\, \,
The work is supported in part by
National Natural Science Foundation of China (grant Nos. 11271138, 11531004), China Scholoarship Council and Simons Foundation grant No. 198129.


\bigskip
\begin{thebibliography}{99}

\bibitem{GaH} A. Gabriel, B. Hiesmayr and M. Huber, Quant. Inf. Comput. \textbf{10}, 829 (2010).

\bibitem{HuS} M. Huber and R. Sengupta, Phys. Rev. Lett. \textbf{113}, 100501 (2014).

\bibitem{Huber} M. Huber, F. Mintert, A. Gabriel and B. C. Hiesmayr, Phys. Rev. Lett. \textbf{104}, 210501 (2010).

\bibitem{WuK} J.Y. Wu, H. Kampermann, D. Bru\ss, C. Klockl, and M.
Huber, Phys. Rev. A \textbf{86}, 022319 (2012).

\bibitem{HuP} M. Huber, M. Perarnau-Llobet, J.I. de Vicente, Phys.
Rev. A \textbf{88}, 042328 (2013).

\bibitem{SpV} J. Sperling, W. Vogel, Phys. Rev. Lett. \textbf{111}, 110503
(2013).


\bibitem{JuM} B. Jungnitsch, T. Moroder, and O. G¡§uhne, Phys. Rev.
Lett. \textbf{106}, 190502 (2011). 
%

\bibitem{MaL} M. Markiewicz, W. Laskowski, T. Paterek, and M.
Z¨B ukowski Phys. Rev. A \textbf{87}, 034301 (2013).

\bibitem{dVH} J. de Vicente and M. Huber, Phys. Rev. A \textbf{84}, 062306 (2011).

\bibitem{MaC} Z.H. Ma, Z.H. Chen, J.L. Chen, C. Spengler, A. Gabriel,
and M. Huber, Phys. Rev. A \textbf{83}, 062325(2011). 

\bibitem{ChM} Z.H. Chen, Z.H. Ma, J.L. Chen, and S. Severini, Phys.
Rev. A \textbf{85}, 062320 (2012). 


\bibitem{HoG} Y. Hong, T. Gao, and F.L. Yan, Phys. Rev. A \textbf{86}, 062323
(2012). 

\bibitem{GaY} T. Gao,. F.L. Yan, and S.J. van Enk, Phys. Rev. Lett.
\textbf{112}, 180501 (2014). 

\bibitem{LiF} M. Li, S.-M. Fei, X. Li-Jost and H. Fan, Phys. Rev. A \textbf{92}, 062338 (2015).

\bibitem{BaG} J.D. Bancal, N. Gisin, Y.C. Liang, and S. Pironio, Phys.
Rev. Lett. \textbf{106}, 250404 (2011). 

\bibitem{Acin} A. Ac\'{\i}n, D. Bru{\ss}, M. Lewenstein and A. Sanpera, Phys. Rev. Lett. \textbf{87}, 040401 (2001).

\bibitem{Heisenberg} W. Heisenberg, Z. Phys. \textbf{43}, 172 (1927).

\bibitem{Robertson} H. P. Robertson, Phys. Rev. \textbf{34}, 163 (1929).

\bibitem{XJLF} Y. Xiao, N. Jing, X. Li-Jost, and S.-M. Fei, {\em Weighted uncertainty relations}, Sci. Rep., in press (2016).
(arXiv: 1603.01004)

\bibitem{Hirschman} I. I. Hirschman, Amer. J. Math. \textbf{79}, 152 (1957).

\bibitem{Deutsch} D. Deutsch, Phys. Rev. Lett. \textbf{50}, 631 (1983).

\bibitem{Maassen} H. Maassen and J. B. M. Uffink, Phys. Rev. Lett. \textbf{60}, 1103 (1988).

\bibitem{Coles} P. J. Coles and M. Piani, Phys. Rev. A. \textbf{89}, 022112 (2014).


\bibitem{Wehner} S. Wehner and A. Winter, New J. Phys. \textbf{12}, 025009 (2010).

\bibitem{Blankenbecler} R. Blankenbecler and M. H. Partovi, Phys. Rev. Lett. \textbf{54}, 373 (1985).

\bibitem{Damgaard} I. Damgaard, S. Fehr, L. Salvail, and C. Schaffner, \textbf{26} (2005), arXiv: 0508222 [quant-ph].

\bibitem{DiVincenzo} D. P. DiVincenzo, M. Horodecki, D. W. Leung, J. A. Smolin, and B. M. Terhal, Phys. Rev. Lett. \textbf{92}, 067902 (2004).

\bibitem{Oppenheim} J. Oppenheim and S. Werner, Science \textbf{330}, 1072 (2010).

\bibitem{Goehne} O. G\"{u}hne, Phys. Rev. Lett. \textbf{92}, 117903 (2004).

\bibitem{Loock} S. L. Braunstein and P. Van Loock, Rev. Mod. Phys. \textbf{77}, 513 (2005).

\bibitem{Coless} Coles, P. J., Berta, M., Tomamichel, M. and Wehner, S. 
arXiv: 1511.04857 (2015).

\bibitem{Partovi} M. H. Partovi, Phys. Rev. A. \textbf{84}, 052117 (2011).

\bibitem{Friedland} S. Friedland, V. Gheorghiu, G. Gour, Phys. Rev. Lett. \textbf{111}, 230401 (2013).

\bibitem{Puchala} Z. Puchala, L. Rudnicki, K. Zyczkowski, J. Phys. A \textbf{46}, 272002 (2013).

\bibitem{Marjorization} M. Albert, W. O. Ingram, and B. C. Arnold, \textit{Inequalities: Theory of Majorization and Its Applications}, 2nd. ed.,
Springer Ser. in Stat., Springer, 2011.

\bibitem{Hossein} M. H. Partovi, Phys. Rev. A. \textbf{86}, 022309 (2012).

\bibitem{Bourennane} M. Bourennane, M. Eibl, C. Kurtsiefer, S. Gaertner, H. Weinfurter, O. G\"{u}hne, P. Hyllus, D. Bru{\ss}, M. Lewenstein and A. Sanpera, Phys. Rev. Lett. \textbf{92}, 087902 (2004).

\bibitem{Schur} I. Schur, Theorie Sitzungsber Berlin. Math. Gesellschaft \textbf{22}, 9 (1923).

\bibitem{Ando} T. Ando, Linear Alg. Appl. \textbf{118}, 163 (1989). 

\bibitem{Igor} F. Liese and I. Vajda, IEEE Trans. Inf. Theory \textbf{52}, 881731 (2006).

\bibitem{Tsallis} J. Havrda and F. Charvat, Kybernetica \textbf{3}, 30 (1967).

\bibitem{Rudnicki} {\L}. Rudnicki, Z. Pucha{\l}a and k. \.{Z}yczkowski, Phys. Rev. A. \textbf{89}, 052115 (2014).

\end{thebibliography}
\end{document}